\documentclass[twocolumn,showpacs,showkeys]{revtex4}

\begin{document}

% Use the \preprint command to place your local institutional report
% number in the upper righthand corner of the title page in preprint mode.
%\preprint{}

%Title of paper
\title{Dynamically broken Anti-de Sitter action for gravity}

% Explanatory text should go in the []'s, actual e-mail
% address or url should go in the {}'s for \email and \homepage.
% Please use the appropriate macro foreach each type of information

\author{Romualdo Tresguerres}
\email[]{romualdotresguerres@yahoo.es}
%\homepage[]{Your web page}
%\thanks{}
%\altaffiliation{}
\affiliation{Instituto de Matem\'aticas y F\'isica Fundamental\\
Consejo Superior de Investigaciones Cient\'ificas\\ Serrano 113
bis, 28006 Madrid, SPAIN}

\date{\today}

\begin{abstract}
Due to a suitable Higgs mechanism, a standard Anti-de Sitter gauge theory becomes spontaneously broken. The resulting Lorentz invariant gravitational action includes the Hilbert-Einstein term of ordinary Einstein-Cartan gravity with cosmological constant, plus contributions quadratic in curvature and torsion, and a scalar Higgs sector.
\end{abstract}

% insert suggested PACS numbers in braces on next line
\pacs{04.50.Kd, 11.15.Ex}
% insert suggested keywords - APS authors don't need to do this
\keywords{Nonlinear realizations, symmetry breaking, gauge gravity, spacetime Higgs multiplet, broken gravitational action, extended Standard Model.}
\maketitle

% body of paper here - Use proper section commands
% References should be done using the \cite, \ref, and \label commands

\section{Introduction}
% Put \label in argument of \section for cross-referencing
%\section{\label{}}

Spontaneous symmetry breaking finds a natural formulation in terms of nonlinear realizations (NLR's) of gauge groups \cite{Cho:1978ss} \cite{Tiemblo:1996gq} \cite{Sardanashvily:2005mf} \cite{Kirsch:2005st}. Given the gauge theory of a group $G$, a suitable NLR allows to reduce the manifest symmetry of the action to that of an arbitrary subgroup $H\subset G$, thus preparing the breakdown from $G$ to the residual symmetry $H$. However, NLR's by themselves do not imply the breaking of the original symmetry $G$. Different realizations of $G$ with alternative choices of the explicit symmetry subgroup $H$ are possible, all of them having in common their invariance under the underlying full gauge group $G$. For instance, as shown by us elsewhere \cite{Lopez-Pinto:1997aw} \cite{Tiemblo:2005js}, the entire (unbroken) Poincar\'e symmetry can be realized at convenience either with $H=$Lorentz or with $H=SO(3)$, being the particular choices mutually reversible at any time. The effective breaking of the symmetry requires a dynamical mechanism which is not automatically provided by NLR's. In the present paper we study the symmetry breaking of an Anti-de Sitter gauge theory, using a modification of the approach due to Stelle and West \cite{Stelle:1980aj} \cite{Leclerc:2005qc}, who first adapted the Higgs mechanism to spacetime gauge symmetries. Our original action (\ref{totLag1}) includes an ordinary kinetic term built from the gauge potentials, plus Higgs contributions involving a Higgs-type multiplet, in quite a close analogy with the standard breaking procedure of the Weinberg-Salam model \cite{Abers:1973qs}. The lastly deduced broken action (\ref{totLag2}) is Lorentz invariant. It resembles a nonlinearly realized action, with the difference that in it the original $G$ symmetry is lost.

\section{de Sitter and Anti-de Sitter group algebras}
%\subsection{}

Let us start introducing the generators $L_{AB}$ of the de Sitter group $O(4\,,1)$ (resp. the Anti-de Sitter group $O(3\,,2)$\,), antisymmetric in the indices $A$, $B$, and satisfying the commutation relations
\begin{equation}
[\,L_{_{AB}}\,, L_{_{MN}}]=-i\,( o_{_{A[M}}L_{_{N]B}} - o_{_{B[M}}L_{_{N ]A}})\,.\label{comrel0}
\end{equation}
The Lie algebras of both groups, with the common form (\ref{comrel0}), differ in the characteristic five-dimensional constant metric involved in each case, namely
\begin{equation}
o_{_{AB}}=diag (-\,,+\,,+\,,+\,,\lambda\,)\quad {\rm with}\quad\lambda =\pm 1\,,\label{metric0}
\end{equation}
where $\lambda =1$ corresponds to the de Sitter, and $\lambda =-1$ to the Anti-de Sitter group. In the fiber bundle approach to be followed by us, where spacetime symmetries are treated in analogy to the internal groups of ordinary gauge theories, the five internal dimensions of the fundamental representation space of the de Sitter group (equipped with the metric (\ref{metric0})) are not correlated to the dimensionality of the (spacetime) base space. Nevertheless, in order to facilitate the nonlinear realization of the (Anti)-de Sitter group $G$ with the Lorentz group as the $H$ subgroup, we proceed in analogy to the group contraction formalism \cite{Inonu:1953sp} \cite{Gursey:1964} to decompose (\ref{metric0}) as
\begin{equation}
o_{\alpha\beta}=diag (-\,,+\,,+\,,+\,)\,,\qquad o_{55}=\lambda\,,\label{metric1}
\end{equation}
and to define the pseudotranslational generators
\begin{equation}
\Pi _\mu :={2\over l}\,L_{\mu 5}\,,\label{pseudotrans}
\end{equation}
with $l$ as a constant with dimensionality of length. By replacing (\ref{metric1}) and (\ref{pseudotrans}) into (\ref{comrel0}), we get the commutation relations
\begin{equation}
[\,L_{\alpha\beta}\,, L_{\mu\nu}]=-i\,( o_{\alpha [\mu}L_{\nu ]\beta} - o_{\beta [\mu}L_{\nu ]\alpha})\,,\label{comrel1}
\end{equation}
\begin{equation}
[\,L_{\alpha\beta}\,,\Pi _\mu ]=i\,o_{\mu [\alpha}\Pi _{\beta ]}\,,\label{comrel2}
\end{equation}
\begin{equation}
[\,\Pi _\mu\,,\Pi _\nu ]={{2i\lambda}\over{l^2}}\,L_{\mu\nu}\,,\label{comrel3}
\end{equation}
differing from those of the Poincar\'e group merely in the occurrence of (\ref{comrel3}) instead of $[\,\Pi _\mu\,,\Pi _\nu ]= 0$, so that the operators $\Pi _\mu$ can be brought to mimetize the behavior of translational generators by taking the limit $l\rightarrow\infty$ considered in the context of group contractions. Here we don't pay attention to this limit.

\section{\bf{Nonlinear approach}}

The mathematical structure of gauge theories is that of fiber bundles \cite{Daniel:1979ez} \cite{Eguchi:1980jx}, with fibers diffeomorphic to the structure group $G$ attached to each point of the base space $M$. A fiber bundle interpretation of nonlinear gauge realizations of groups was proposed by us in \cite{Tresguerres:2002uh}. We choose the base manifold from the beginning to be four-dimensional in view of its future identification as physical spacetime.

Let us summarize the essentials of the nonlinear approach. Given a group $G$ with a subgroup $H$, we define an action of $G$ on the quotient space $G/H$ as follows. At any point $x\in M$, the group element $g(x)\in G$, acting on the left on cosets $\sigma _\xi\in G/H$ labeled by continuous parameters $\xi (x)$, yields
\begin{equation}
L_g\circ\sigma _\xi (x)=R_h\circ\sigma _{\xi '}(x)\,,\label{nonlintrans2}
\end{equation}
where the r.h.s. expresses the right action of $h(\xi\,,g)\in H$ on a transformed coset $\sigma _{\xi '}\in G/H$ with $\xi '\,(\xi\,,g)$.

Provided that linear representations $\rho (h)$ of the subgroup elements $h\in H$ exist, (\ref{nonlintrans2}) induces an action of $g\in G$ on fields $\psi$ of the representation spaces of $H$ as
\begin{equation}
\psi '\,=\rho (h(\xi\,,g))\psi\,.\label{nonlinfieldtransf}
\end{equation}
Despite of being (\ref{nonlinfieldtransf}) formally a linear $H$-transformation, the nonlinear action of the full group $G$ enters it through the dependence $h(\xi\,,g)$ as derived from (\ref{nonlintrans2}).

A capital feature of NLR's related with the previous one is the occurrence of $G$-nonlinear connections $\Gamma _{_M}$ formally transforming as $H$-gauge potentials, their relation to the standard $G$-linear connections $A_{_M}$ being given by
\begin{equation}
\Gamma _{_M}=\sigma _\xi ^{-1}(\,d + A_{_M} )\,\sigma _\xi\,.\label{compar4}
\end{equation}
Readers interested in details are referred to the literature \cite{Tresguerres:2002uh} \cite{Coleman:1969sm} \cite{Tresguerres:2000qn} \cite{Tiemblo:2005sx}.

Let us now apply the general formalism to the case of $G=$(Anti)-de Sitter and $H=$Lorentz. We rewrite (\ref{nonlintrans2}) in the form $g\cdot\sigma _\xi =\sigma _{\xi '}\cdot h\,$, see \cite{Coleman:1969sm} \cite{Tiemblo:2005sx}, and replace the infinitesimal group elements $g\approx I +i\,\beta ^{AB} L_{AB} = I+i\,\beta ^{\alpha\beta} L_{\alpha\beta}+i\,\epsilon ^\mu\Pi _\mu $ of the (Anti)-de Sitter group (with $\epsilon ^\mu :=l\beta ^{\mu 5}\,$), and $h\approx I+i\,\mu ^{\alpha\beta} L_{\alpha\beta}$ of the Lorentz group. On the other hand, we parametrize the cosets as $\sigma _\xi =e^{-i\,\xi ^\mu\Pi _\mu}$ with finite pseudotranslational parameters $\xi ^\mu (x)$, and $\sigma _{\xi '} =e^{-i\,{\xi'}^\mu\Pi _\mu}$ with ${\xi '}^\mu\approx\xi ^\mu +\delta\xi ^\mu$. Then, with the help of the Hausdorff-Campbell formula, we get for $\lambda =1$ the infinitesimal variation
\begin{equation}
\delta\xi ^\mu =-\xi ^\nu \beta _\nu{}^\mu -\epsilon ^\nu \left[ (\delta _\nu ^\mu -{{\xi _\nu\xi ^\mu}\over{\xi ^2}}){(|\xi |/l)\over{\tan (|\xi |/l)}}+{{\xi _\nu\xi ^\mu}\over{\xi ^2}}\right]\label{varxilamb+}
\end{equation}
of the pseudotranslational coset parameters, with $|\xi |:=\sqrt{ o_{\alpha\beta}\,\xi ^\alpha\xi ^\beta}$. (And for $\lambda =-1$, an analogous expression with $\tan (|\xi |/l)$ replaced with  $\tanh (|\xi |/l)\,$). On the other hand, the infinitesimal nonlinear Lorentz parameters entering (\ref{nonlinfieldtransf}) as much as the gauge transformations (\ref{varlorconn}) and (\ref{vartetrad}) of the components of (\ref{compar4}), are found to be
\begin{equation}
\mu ^{\alpha\beta} =\beta ^{\alpha\beta}+{{2\lambda\,\sin (|\xi |/l)}\over{l\,|\xi |\,(1+\cos (|\xi |/l))}}\epsilon ^{[\alpha }\xi ^{\beta ]}\label{Hparamredefbis}
\end{equation}
for $\lambda =1$. (An analogous result holds for $\lambda =-1$ with the trigonometrical functions replaced by the corresponding hyperbolic ones.) Next we build the nonlinear connection (\ref{compar4}) of the (Anti)-de Sitter group starting from the standard linear connection
\begin{equation}
A_{_M} =-i\,\Gamma ^{AB} L_{AB}\,,\label{lindeSitter1}
\end{equation}
transforming as
\begin{equation}
\delta\Gamma ^{AB} ={\buildrel (5)\over{D}}\beta ^{AB} :=d\beta ^{AB}+\Gamma _C{}^A\beta ^{CB}+\Gamma
_C{}^B\beta ^{AC}\,.\label{vartotconnect}
\end{equation}
(From now on, we denote with ${\buildrel (5)\over{D}}$ the covariant derivative built with the linear connection $\Gamma ^{AB}$, in order to distinguish it from the nonlinear Lorentz covariant derivative $\hat{D}$, see for instance (\ref{varlorconn}) below.) For calculational convenience, we make use of (\ref{pseudotrans}) and of definition ${\buildrel (T)\over{\Gamma ^\mu}}:=l\,\Gamma ^{\mu 5}\,$ to rewrite (\ref{lindeSitter1}) as
\begin{equation}
A_{_M} =-i\,\Gamma ^{\alpha\beta} L_{\alpha\beta} -i\,{\buildrel (T)\over{\Gamma ^\mu}}\Pi _\mu\,,\label{lindeSitter2}
\end{equation}
where neither $\Gamma ^{\alpha\beta}$ is a true Lorentz connection nor ${\buildrel (T)\over{\Gamma ^\mu}}$ behaves as a Lorentz coframe, as easily read out from (\ref{vartotconnect}). However, by replacing (\ref{lindeSitter2}) in (\ref{compar4}) together with the cosets $\sigma _\xi =e^{-i\,\xi ^\mu\Pi _\mu}$, we find the nonlinear connection
\begin{equation}
\Gamma _{_M}=-i\,\hat{\Gamma}^{\alpha\beta}L_{\alpha\beta} -i\,\vartheta ^\mu\Pi _\mu\,,\label{nlindeSitter}
\end{equation}
resembling (\ref{lindeSitter2}), but with modified components transforming differently, see (\ref{varlorconn}), (\ref{vartetrad}). The search for a simple explicit notation for the components of (\ref{nlindeSitter}) is facilitated by introducing the matrix
\begin{equation}
M_{_{AB}}:=(\sigma _\xi )_{_{AB}}\,,\label{bmatrix}
\end{equation}
in the fundamental five-dimensional representation of the (Anti)-de Sitter group. (The explicit form of (\ref{bmatrix}) is given in Appendix A.) In terms of (\ref{bmatrix}), the nonlinear connection (\ref{compar4}) with (\ref{lindeSitter1}) yields
\begin{equation}
\Gamma _{_M}=-i\,(M^{-1})^{_B}{}_{_C}(\,dM^{_{CA}}+\Gamma _{_D}{}^{_C}M^{_{DA}})L_{_{AB}}\,,\label{nlindeSitterbis}
\end{equation}
to be compared with (\ref{nlindeSitter}) to get
\begin{equation}
\hat{\Gamma}^{\alpha\beta} = \Gamma ^{\alpha\beta} - M^{A [\,\alpha}\,{\buildrel (5)\over{D}}(M^{-1})^{\,\beta\,]}{}_A\,,\label{nonlinlorconlamb}
\end{equation}
and --with a simplification allowed by the form (\ref{bmatrixcomp1down})--(\ref{bmatrixcomp4down}) of (\ref{bmatrix})--
\begin{equation}
\vartheta ^\mu = {\buildrel (T)\over{\Gamma ^\mu}} +l\, M^{A5}{\buildrel (5)\over{D}}(M^{-1})^\mu{}_A\,.\label{tetradlamb1}
\end{equation}
The crucial point concerning (\ref{nonlinlorconlamb}) and (\ref{tetradlamb1}) is that their manifest transformation properties are Lorentz ones involving the parameters (\ref{Hparamredefbis}). Actually, for the Lorentz connection (\ref{nonlinlorconlamb}) we can calculate (see \cite{Tresguerres:2002uh})
\begin{equation}
\delta\hat{\Gamma}^{\alpha\beta} =\hat{D}\mu ^{\alpha\beta}\,,\label{varlorconn}
\end{equation}
where $\hat{D}$ is the Lorentz covariant differential built with (\ref{nonlinlorconlamb}) itself. In addition, it is a general feature of NLR's that those components of the nonlinear connection which are associated to group generators different from the $H$ ones become $H$-tensors. We find in particular the pseudotranslational components (\ref{tetradlamb1}) to obey the gauge transformations
\begin{equation}
\delta\vartheta ^\mu =-\vartheta ^\nu \mu _\nu{}^\mu\,,\label{vartetrad}
\end{equation}
characteristic for a Lorentz covector, thus making it possible to interpret such 1-forms $\vartheta ^\mu$ as tetrads \cite{Lopez-Pinto:1997aw} \cite{Tiemblo:2005js} \cite{Tresguerres:2002uh} \cite{Tresguerres:2000qn} \cite{Tiemblo:2005sx}. One can write them down as $\vartheta ^\mu = dx^i e_i{}^\mu$ with $e_i{}^\mu$ being $4\times 4$ matrices, since the explicit group index runs from $0$ to $3$, and so does the hidden coordinate index referring to the base space $M$, previously chosen by us to be $4$-dimensional. In Appendix B, from the tetrads we build the eta-basis providing a convenient notation for what follows.

\section{Field strengths and Yang-Mills (Anti)-de Sitter action}

Let us handle the (Anti)-de Sitter group as an ordinary gauge group. In order to build its pure Yang-Mills action we introduce the linear covariant differential operator
\begin{equation}
{\buildrel (\,5)\over D}:=d+ A_{_M}\,,\label{lcovdiff1}
\end{equation}
with $A_{_M}$ as given by (\ref{lindeSitter1}). Commuting (\ref{lcovdiff1}) as
\begin{equation}
{\buildrel (\,5)\over D}\wedge {\buildrel (\,5)\over D} =d A_{_M} +A_{_M}\wedge A_{_M} =-i\,F^{AB}L_{AB}\,,\label{lcurv1}
\end{equation}
we find the field strength 2-form (or {\it curvature}) defined as
\begin{equation}
F^{AB} := d\Gamma ^{AB} +\Gamma _C{}^B\wedge\Gamma ^{AC}\,.\label{lcurv1def}
\end{equation}
The corresponding standard linear (Anti)-de Sitter Yang-Mills kinetic term reads
\begin{equation}
L^{YM}=-{1\over 4}\,F^{AB}\wedge\,{}^*F_{BA}\,.\label{YMLag1}
\end{equation}
Let us now rewrite the Lagrangian density (\ref{YMLag1}) in terms of the nonlinear variables introduced in previous Section. In analogy to (\ref{lcovdiff1}) we define the nonlinear covariant differential operator
\begin{equation}
{\buildrel {NL}\over D}:=d+\Gamma _{_M}\,,\label{nlcovdiff1}
\end{equation}
involving the nonlinear connection (\ref{compar4}). In parallel to (\ref{lcurv1}) we get
\begin{eqnarray}
{\buildrel {NL}\over D}\wedge {\buildrel {NL}\over D} &=& d\,\Gamma _{_M} +\Gamma _{_M}\wedge \Gamma _{_M}\nonumber \\
&=:&-i\,{\buildrel {NL}\over {F^{AB}}} L_{AB}\nonumber \\
&=& -i\,{\buildrel {NL}\over {F^{\alpha\beta}}} L_{\alpha\beta}  -i\,l {\buildrel {NL}\over {F^{\mu 5}}} \Pi _\mu \,,\label{lcurv2}
\end{eqnarray}
where the components of the nonlinear field strength ${\buildrel {NL}\over {F^{AB}}}$ can be explicitly found by calculating from (\ref{nlindeSitter})
\begin{equation}
d\,\Gamma _{_M} +\Gamma _{_M}\wedge \Gamma _{_M} = -i\,\left(\,\hat{R}^{\alpha\beta}+ {\lambda\over{l^2}}\,\vartheta ^\alpha\wedge\vartheta ^\beta\,\right) L_{\alpha\beta} -i\,T ^\mu\,\Pi _\mu\,,\label{lcurv3bis}
\end{equation}
so that comparing (\ref{lcurv2}) with (\ref{lcurv3bis}) one gets
\begin{eqnarray}
{\buildrel {NL}\over {F^{\alpha\beta}}} &=& \hat{R}^{\alpha\beta}+ {\lambda\over{l^2}}\,\vartheta ^\alpha\wedge\vartheta ^\beta \,,\label{nlfieldstrengthcomp1}\\
{\buildrel {NL}\over {F^{\mu 5}}} &=& {1\over l}\,T ^\mu \,,\label{nlfieldstrengthcomp2}
\end{eqnarray}
where we recognize well known {\it geometrical} objects such as the ordinary Lorentz curvature 2-form
\begin{equation}
\hat{R}_\alpha{}^\beta := d\hat{\Gamma}_\alpha{}^\beta
+\hat{\Gamma}_\gamma{}^\beta\wedge\hat{\Gamma}_\alpha{}^\gamma
\,,\label{curv}
\end{equation}
and the pseudotranslational field strength
\begin{equation}
T ^\mu :=\hat{D}\vartheta ^\mu =d\vartheta ^\mu
+\hat{\Gamma}_\nu{}^\mu\wedge\vartheta ^\nu\,,\label{torsion}
\end{equation}
the latter having the same form and transformation properties as the ordinary torsion of Poincar\'e Gauge Theories. (We use the exterior calculus notation as fixed by \cite{Hehl:1995ue}.) From (\ref{compar4}) we also find
\begin{equation}
d\,\Gamma _{_M} +\Gamma _{_M}\wedge \Gamma _{_M} =\sigma _\xi ^{-1}(\,d A_{_M} +A_{_M}\wedge A_{_M} )\,\sigma _\xi\,.\label{lcurv2bis}
\end{equation}
Using (\ref{nlindeSitterbis}) and the notation introduced in (\ref{lcurv1}) and (\ref{lcurv2}), the relation (\ref{lcurv2bis}) can alternatively be expressed as
\begin{equation}
{\buildrel {NL}\over {F^{AB}}} =(M^{-1})^A{}_C F^{CD}M_D{}^B\,.\label{curvrel}
\end{equation}
Let us now apply these results to (\ref{YMLag1}). From (\ref{curvrel}) follows the equality
\begin{equation}
F^{AB}\wedge\,{}^*F_{BA} = {\buildrel {NL}\over {F^{AB}}}\wedge\,{}^*{\buildrel {NL}\over {F_{BA}}}\,,\label{kin1}
\end{equation}
so that (\ref{YMLag1}) admits an invariant reformulation into
\begin{eqnarray}
L^{YM}&=&-{1\over 4}\,F^{AB}\wedge\,^*F_{BA}= -{1\over 4}\,{\buildrel {NL}\over {F^{AB}}}\wedge\,{}^*{\buildrel {NL}\over {F_{BA}}}\nonumber \\
&=& -{1\over 4}\,\left(\,{\buildrel {NL}\over {F^{\alpha\beta}}}\wedge\,{}^*{\buildrel {NL}\over {F_{\beta\alpha}}} -2\,{\buildrel {NL}\over {F^{\mu 5}}}\wedge\,{}^*{\buildrel {NL}\over {F_{\mu 5}}}\,\right)\,.\label{YMLag2}
\end{eqnarray}
Making use of (\ref{nlfieldstrengthcomp1}) and (\ref{nlfieldstrengthcomp2}), we finally arrive at
\begin{eqnarray}
L^{YM} &=&{{\lambda}\over{2l^2}}\,\hat{R}^{\alpha\beta}\wedge\eta _{\alpha\beta} +{3\over{l^4}}\,\eta\nonumber\\
&+&{1\over 4}\,\hat{R}^{\alpha\beta}\wedge\,^*\hat{R}_{\alpha\beta} +{1\over{2l^2}}\,T ^\mu\wedge\,^*T _\mu\,,\label{YMLag3}
\end{eqnarray}
(see Appendix B for the eta-basis notation), where the first term is (but for the sign) an ordinary Einstein-Cartan Lagrange density, see for instance \cite{Hehl:1979xk} \cite{Gronwald:1995em}. The remaining contributions are a cosmological constant term (with a troublesome prefixed value related to that of the Newton constant), and further terms quadratic in curvature and torsion respectively.

\section{Defining an (Anti)-de Sitter Higgs-type multiplet}

From the way we followed to deduce (\ref{YMLag3}), it is clear that this Lagrangian density is a mere reformulation of (\ref{YMLag1}) in terms of the nonlinear quantities (\ref{nonlinlorconlamb}), (\ref{tetradlamb1}). Accordingly, the underlying (Anti)-de Sitter invariance survives despite the formal Lorentz character of the transformations (\ref{varlorconn}), (\ref{vartetrad}) obeyed by the new variables. The effective breakdown of the original symmetry to the residual Lorentz invariance actually requires an additional dynamical breaking procedure. Here we present an adaptation of the standard Higgs mechanism.

The latter rests on the introduction of a certain Higgs multiplet \cite{Abers:1973qs}. In analogy to it, we introduce a five-vector corresponding to the fundamental representation of the (Anti)-de Sitter group, parametrized as the product
\begin{equation}
y^{_A}:=M^{_A}{}_{_B}(u\,\delta _5^{_B})\label{5vector}
\end{equation}
of the matrix (\ref{bmatrix}) (see Appendix A) times a multiplet $(u\,\delta _5^{_B})$ whose only nonvanishing component is taken to be a scalar field $u$. Such singlet introduced by hand is the only element of the present theory that is not deductively provided by the NLR of the gauge group $G$. But it is indeed necessary, besides the four pseudotranslational coset fields $\xi ^\mu$ entering the matrix (\ref{bmatrix}), to complete the five degrees of freedom of the vector (\ref{5vector}). Taking (\ref{bmatrixcomp1down})--(\ref{bmatrixcomp4down}) into account, the components of (\ref{5vector}) are found to be, for the choice $\lambda =1$ \begin{equation}
y^\mu =-u\,{{\xi ^\mu}\over{|\xi |}}\,\sin (|\xi |/l)\,,\quad y^5 =u\,\cos (|\xi |/l)\,,\label{5vectlamb+}
\end{equation}
and respectively for $\lambda =-1$
\begin{equation}
y^\mu =u\,{{\xi ^\mu}\over{|\xi |}}\,\sinh (|\xi |/l)\,,\quad y^5 =u\,\cosh (|\xi |/l)\,,\label{5vectlamb-}
\end{equation}
so that in both cases
\begin{equation}
y_{_A} y^{_A} =y_\mu y^\mu +\lambda\,(y^5)^2 =\lambda\,u^2\,.\label{uinvartlamb}
\end{equation}
Provided $\delta u =0$ as corresponding to a scalar field, from (\ref{5vectlamb+}) with (\ref{varxilamb+}) (and analogously for the $\lambda =-1\,$ case), we find
\begin{equation}
\delta y^A =-y^B\beta _B{}^A\,,\label{vary1}
\end{equation}
valid for both, $\lambda =\pm 1$, so that (\ref{5vector}) is a true O(4\,,1) (resp. $O(3\,,2)$ ) vector.

The variables $\xi ^\mu$ entering the matrix (\ref{bmatrixbis}) used in (\ref{5vector}) are Goldstone fields, having as a main feature to be eliminable from an (Anti)-de Sitter invariant action by means of a gauge-like redefinition of the remaining fields and connections. The so called {\it unitary gauge fixing} \cite{Abers:1973qs} handles the Goldstone matrix (\ref{bmatrixbis}) as if it were a gauge transformation matrix, implicating it on the one hand in the {\it gauging away} of the Goldstone degrees of freedom from (\ref{5vector}), yielding
\begin{equation}
\hat{y}^{_A} =u\,\delta _5^{_A}\,,\label{unitgaugey}
\end{equation}
and on the other hand in the simultaneous field rearrangement giving rise to the nonlinear variables (\ref{nonlinlorconlamb}) and (\ref{tetradlamb1}). Actually, fixing the {\it unitary gauge} is equivalent to introduce a NLR.

Certainly, NLR's can be performed independently from the breaking of the symmetry, as already shown by the reformulation of (\ref{YMLag1}) as (\ref{YMLag3}). Instead, the breakdown of the symmetry from $G$ to $H$ is inseparable from the corresponding NLR, necessarily giving rise to a {\it unitary gauge fixing}. Even if not declared, the parametrization (\ref{5vector}) of the Higgs multiplet (\,favoring a particular symmetry breakdown) implicitly presupposes a certain expected NLR. Given the initial gauge group $G$, the role of the subgroup $H$ is singled out in advance by the choice of the Goldstone matrix (\ref{bmatrixbis}), containing coset parameter fields $\xi ^\mu$ taken {\it ad hoc} to belong to $G/H$. Accordingly, when later the effective symmetry breaking occurs, the subgroup $H$ emerges --not surprisingly-- as the natural residual invariance.

\section{The Higgs mechanism}

In our approach, the spontaneous breaking of the original symmetry is due to a special treatment of the dynamics of the Higgs vector field (\ref{5vector}), similar to that of the Standard Model. We consider the explicitly (Anti)-de Sitter invariant Lagrangian density
\begin{equation}
L^{Higgs}= k_0\,{\buildrel (5)\over{D}}\,y^A\wedge\,^*{\buildrel (5)\over{D}}\,y_A + k_1\,^*(y^A y_A) +k_2\,^*\bigl[\,(y^A y_A)^2\,\bigr]\,,\label{HiggsLag1}
\end{equation}
including a kinetic and a mass term plus a quartic selfinteraction contribution. (The Hodge dual star operators refer to the four-dimensional base space. In view of (\ref{vol}), they can be replaced by the volume element built with the tetrads, thus being explicitly Lorentz, and at the same time nonlinearly (Anti)-de Sitter, invariant.) The constants in (\ref{HiggsLag1}) will be fixed at convenience in due course. For our purpose, it is appropriate to transform the $y^{_A}$-kinetic term as follows. With the help of (\ref{bmatrixcomp1down})--(\ref{bmatrixcomp4down}) and (\ref{5vector}), we find for the tetrad (\ref{tetradlamb1}) the alternative form
\begin{equation}
\vartheta ^\mu =-\lambda\,l\,(M^{-1})^\mu{}_A\,{\buildrel (5)\over{D}}({\,\,y^A\over u})\,,\label{tetradlamb2}
\end{equation}
trivially implying
\begin{equation}
{\buildrel (5)\over{D}}({\,\,y^A\over u})=-{{\lambda}\over l}\,M^A{}_\mu\,\vartheta ^\mu\,.\label{ycovder}
\end{equation}
From (\ref{ycovder}) it is easily calculated that the kinetic term in (\ref{HiggsLag1}) simplifies to
\begin{equation}
{\buildrel (5)\over{D}}\,y^A\wedge\,^*{\buildrel (5)\over{D}}\,y_A =\lambda\,du\wedge\,^*du +{{u^2}\over{l^2}}\,\vartheta ^\mu\wedge\,^*\vartheta _\mu\,.\label{Higgskin}
\end{equation}
In (\ref{Higgskin}) the Goldstone fields entering $y^{_A}$ have been entirely absorbed into the tetrads, see (\ref{tetradlamb1}). According to (\ref{feat2}), the last term in (\ref{Higgskin}) can be further brought to $\vartheta ^\mu\wedge\,^*\vartheta _\mu =4\,\eta\,$, so that the whole $y^{_A}$-kinetic term reduces to a kinetic plus a mass contribution for the singlet (or scalar Higgs field) $u\,$. The original Higgs Lagrangian density (\ref{HiggsLag1}), with the replacement of (\ref{Higgskin}) and taking (\ref{uinvartlamb}) into account, converts into
\begin{equation}
L^{Higgs}=\lambda\,k_0\, du\wedge\,^*du +V(u)\,\eta\,,\label{HiggsLag2}
\end{equation}
where we introduced the potential (including mass terms) defined as
\begin{equation}
V(u):=\Bigl(\,{{4k_0}\over{l^2}} +\lambda\,k_1\,\Bigr)\,u^2 + k_2\,u^4\,.\label{Vupotential}
\end{equation}
From now on we limit ourselves to consider $\lambda =-1$ in order to ensure the occurrence of the right sign \cite{Hehl:1979xk} \cite{Gronwald:1995em} in the Einstein-Cartan term in the Lagrangian (\ref{YMLag3}), thus concentrating on the Anti-de Sitter case.

Let us show how the effective spontaneous breaking of the symmetry comes about. The potential (\ref{Vupotential}) is required to have a minimum at $u_0\neq 0$. We find the extremal condition $V'(u_0)=0\,$ to be actually satisfied for the vacuum expectation value
\begin{equation}
u_0 =\sqrt{ -{1\over{2k_2}}\Bigl(\,{{4k_0}\over{l^2}}- k_1\,\Bigr)}\,,\label{extrem}
\end{equation}
while the minimum requirement $V''(u_0)>0\,$ demands
\begin{equation}
k_1 > {{4k_0}\over{l^2}}\,.\label{minim}
\end{equation}
In addition we impose
\begin{equation}
V(u_0)={\Lambda\over{l^2}} -{3\over{l^4}}\,,\label{constval}
\end{equation}
in order to replace the inconvenient value of the cosmological term in (\ref{YMLag3}) by the usual one proportional to the undetermined cosmological constant $\Lambda\,$.

As read out from (\ref{uinvartlamb}), the vacuum expectation value $u_0\neq 0$ is degenerate as being in fact the radius of a complete ring of minima $y_0^{_A}$ of the form (\ref{5vector}), where only $u$ is determined (with its value being given by (\ref{extrem})). The fixing of the {\it unitary gauge} has the meaning of a sort of Anti-de Sitter rotation involving the Goldstone matrix (\ref{bmatrixbis}), as a result of which the vacuum expectation value becomes oriented along one of the vector components, as in (\ref{unitgaugey}). But this choice (indistinguishable from the introduction of a NLR) doesn't yet imply symmetry breaking. It is the field expansion $u=u_0+\varphi\,$ around $u_0$ taken as a ground state --implying the field $\varphi$ to have zero vacuum expectation value-- that introduces a shifting of fields provoking the spontaneous breakdown of the original symmetry. (A residual symmetry under the subgroup $H$ remains for the redefined --nonlinear-- quantities.)

By performing such an expansion of $u$ around ({\ref{extrem}), we find (\ref{HiggsLag2}) (with ({\ref{Vupotential}) and $\lambda =-1\,$) to transform into
\begin{equation}
L^{Higgs}=-k_0\,d\varphi\wedge\,^*d\varphi + k_2\,\Bigl[ -u_0^4 +\bigl(\,2 u_0\,\varphi +\varphi ^2\,\bigr) ^2\,\Bigr]\,\eta \,.\label{HiggsLag3}
\end{equation}
The kinetic and mass terms of (\ref{HiggsLag3}) suggest to take the constant values affecting them to be respectively $k_0 =-1/2$ and $k_2 ={m^2}/{(8 u_0^2)}$. Consistently with conditions (\ref{extrem})--(\ref{constval}), this choice yields the complete set of constant values
\begin{eqnarray}
k_0 &=&-{1\over{2}}\,,\label{kzero}\\
k_1 &=& {{m^2}\over{4}}-{{2}\over{l^2}}\,,\label{kone}\\
k_2 &=&{{l^4 m^4}\over{64\,(3-\Lambda l^2)}}\,,\label{ktwo}
\end{eqnarray}
fixing in addition the ground state (\ref{extrem}) to be
\begin{equation}
u_0 ={{2}\over{l^2 m}}\,\sqrt{2\,(3-\Lambda l^2)}\,.\label{extrembis}
\end{equation}
Replacing these values in (\ref{HiggsLag3}) we get
\begin{eqnarray}
L^{Higgs}&=&{1\over{2}}\,d\varphi\wedge\,^*d\varphi\nonumber \\
&&+\Bigl[\,{\Lambda\over{l^2}} -{3\over{l^4}}+{1\over 2}\,\Bigl(\,m\varphi +{{l^2 m^2\varphi ^2}\over{4\sqrt{2\,(3-\Lambda\,l^2)}}}\,\Bigr) ^2\,\Bigr]\eta\,,\nonumber \\
\label{HiggsLag4}
\end{eqnarray}
whose relevance will become apparent immediately.

Let us put together all the previous results to get an organic view. On the one hand we take as the dynamical starting point of the theory the total linear Lagrangian density built as the sum of (\ref{YMLag1}) and (\ref{HiggsLag1}), with the values of the constants $k_0\,$, $k_1\,$, $k_2\,$ fixed to be (\ref{kzero})--(\ref{ktwo}), that is
\begin{eqnarray}
L^{tot}&=&-{1\over 4}\,F^{AB}\wedge\,^*F_{BA}\nonumber \\
&&-{1\over{2}}\,{\buildrel (5)\over{D}}\,y^A\wedge\,^*{\buildrel (5)\over{D}}\,y_A\nonumber \\
&&+ \Bigl(\,{{m^2}\over{4}}-{{2}\over{l^2}}\,\Bigr)\,\,^*(y^A y_A)\nonumber \\
&&+{{l^4 m^4}\over{64\,(3-\Lambda l^2)}}\,\,^*\bigl[\,(y^A y_A)^2\,\bigr]\,.\label{totLag1}
\end{eqnarray}
After the spontaneous symmetry breaking, (\ref{HiggsLag1}) reduces to (\ref{HiggsLag4}). The latter is expressed in the {\it unitary gauge}, see (\ref{unitgaugey}), corresponding to a NLR with $H=$Lorentz. By consistently rewriting (\ref{YMLag1}) as (\ref{YMLag3}) in terms of the same variables as used in (\ref{HiggsLag4}), it becomes possible to add up in particular the cosmological constant contributions, so that the sum of (\ref{YMLag3}) and (\ref{HiggsLag4}) (\,with $\lambda =-1\,$) yields the broken total Lagrangian
\begin{eqnarray}
L^{tot}=&-&{1\over{2l^2}}\,\hat{R}^{\alpha\beta}\wedge\eta _{\alpha\beta} +{{\Lambda}\over{l^2}}\,\eta\nonumber \\
&+&{1\over 4}\,\hat{R}^{\alpha\beta}\wedge\,^*\hat{R}_{\alpha\beta} +{1\over{2l^2}}\,T ^\mu\wedge\,^*T _\mu\nonumber \\
&+&{1\over{2}}\,d\varphi\wedge\,^*d\varphi +{1\over 2}\,\Bigl(\,m\varphi +{{l^2 m^2\varphi ^2}\over{4\sqrt{2\,(3-\Lambda\,l^2)}}}\,\Bigr) ^2 \eta\,,\nonumber\\
\label{totLag2}
\end{eqnarray}
where in the successive rows we find

\noindent i.-the ordinary Einstein-Cartan Lagrangian density with standard (undetermined) cosmological constant,

\noindent ii.-contributions quadratic in curvature and torsion, and

\noindent iii.-the residual scalar Higgs Lagrangian density with mass and higher order terms.

Despite their similitude, a great difference exists between (\ref{YMLag3}) and (\ref{totLag2}), concerning their respective underlying symmetries. While the former constitutes a mere reformulation, always reversible, of (\ref{YMLag1}) in terms of redefined nonlinear variables, preserving the full (Anti)-de Sitter symmetry, instead in the Lagrangian density (\ref{totLag2}) the original symmetry is irreversibly broken by the Higgs mechanism, so that only the explicit residual Lorentz invariance remains. As a main consequence, the value of the cosmological constant is $\Lambda$ --a value introduced in the last term of (\ref{totLag1})-- instead of $3/l^2$ as in (\ref{YMLag3}).

\section{Conclusions}

We have shown that the modified gravitational Lagrangian density (\ref{totLag2}) derives from the standard Yang-Mills Lagrangian (\ref{YMLag1}) with the help of a symmetry breaking procedure which presents a close analogy with the standard Higgs mechanism. It is worth checking the possible renormalizability of the original (unbroken) total linear Lagrangian ({\ref{totLag1}). (Recall that the {\it unitary gauge}, although making apparent the physical field spectrum, actually constitutes an obstacle to recognize the possible renormalizability of the action.)

On the other hand, by regarding (\ref{totLag1}) as the gravitational contribution to the action of an extended Standard Model, and playing with different choices of constants and signs in ({\ref{HiggsLag1}), one can explore possible combinations (and eventual cancelations) between the standard and the gravitational Higgs scalar sectors.

Let us conclude remarking that the formal resources of NLR's reveal to be useful to deal with both, the unbroken gauge theories of gravity retaining the full $G$ symmetry \cite{Lopez-Pinto:1997aw} \cite{Tiemblo:2005js} \cite{Tresguerres:2000qn} \cite{Tresguerres:2007ih}, as much as the broken ones with residual symmetry $H\subset G$.

% If in two-column mode, this environment will change to single-column
% format so that long equations can be displayed. Use
% sparingly.
%\begin{widetext}
% put long equation here
%\end{widetext}

% If you have acknowledgments, this puts in the proper section head.
%\begin{acknowledgments}
%The author is very grateful to ...
% put your acknowledgments here.
%\end{acknowledgments}

% Specify following sections are appendices. Use \appendix* if there
% only one appendix.
\appendix
\section{Elements of the reducing Goldstone matrix}

In the fundamental five-dimensional representation of the (Anti)-de Sitter algebra, the group generators read
\begin{equation}
(L_{_{AB}})_{_C}{}^{_D} =-i\,o_{_{C[A}}\delta _{_{B]}}^{_D}\,,\label{matrixrep1}
\end{equation}
being in particular the pseudotranslational generators
\begin{equation}
(\Pi _\mu )_{_C}{}^{_D} ={2\over l}\,(L_{\mu 5})_{_C}{}^{_D}=-{{2i}\over l}\,o_{_{C}[\mu}\delta _{5]}^{_D}\,.\label{matrixrep2}
\end{equation}
Making use of ({\ref{matrixrep2}), we calculate the matrix representation ({\ref{bmatrix}) of $\sigma _\xi =e^{-i\,\xi ^\mu\Pi _\mu}$ to be
\begin{eqnarray}
M_{_{AB}}&:=&(\sigma _\xi )_{_{AB}}= (e^{-i\,\xi ^\mu\Pi _\mu})_{_{AB}}\nonumber \\
&=&{\sum _{n=0}^\infty}\,{1\over{n!}}\,\bigl[\,({-i\,\xi ^\mu\Pi _\mu}) ^n\,\bigr] _{_{AB}}\nonumber \\
&=& o_{_{AB}} + ({-i\,\xi ^\mu\Pi _\mu}) _{_{AB}}\nonumber \\
&&\hskip0.2cm +{1\over{2!}}\,({-i\,\xi ^\mu\Pi _\mu}) _{_A}{}^{_C}({-i\,\xi ^\mu\Pi _\mu}) _{_{CB}} +...\nonumber\\
\label{bmatrixbis}
\end{eqnarray}
For $\lambda =1$, the elements of (\ref{bmatrixbis}) read
\begin{eqnarray}
M_{\alpha\beta}&=&(M^{-1})_{\alpha\beta}=o_{\alpha\beta} -(1-\cos (|\xi |/l)){{\xi _\alpha \xi _\beta}\over{(\xi _\gamma \xi ^\gamma})}\,,\hskip1.0cm\label{bmatrixcomp1down}\\
M_{5\alpha}&=&-(M^{-1})_{5\alpha} =\lambda\,{{\xi _\alpha}\over{|\xi |}}\,\sin (|\xi |/l)\,,\label{bmatrixcomp2down}\\
M_{\alpha 5}&=&-(M^{-1})_{\alpha 5} =-\lambda\,{{\xi _\alpha}\over{|\xi |}}\,\sin (|\xi |/l)\,,\label{bmatrixcomp3down}\\
M_{55}&=&\hskip0.2cm (M^{-1})_{55}=\lambda\,\cos (|\xi |/l)\,,\label{bmatrixcomp4down}
\end{eqnarray}
and for $\lambda =-1$ we find analogous values with the trigonometrical functions replaced by hyperbolic ones.

\section{Eta-basis}

Following the exterior calculus notation fixed in \cite{Hehl:1995ue}, we define the Levi-Civita object in terms of the tetrads (\ref{tetradlamb1}) using the Hodge dual star operator $^*$ as
\begin{equation}
\eta ^{\alpha\beta\gamma\delta}:=\,^*(\vartheta
^\alpha\wedge\vartheta ^\beta\wedge\vartheta
^\gamma\wedge\vartheta ^ \delta\,)\,,\label{levicivita}
\end{equation}
and with the help of (\ref{levicivita}) we build the remaining elements of the eta-basis, namely
\begin{eqnarray}
\eta ^{\alpha\beta\gamma}&:=&\,^*(\vartheta ^\alpha\wedge\vartheta ^\beta\wedge\vartheta ^\gamma\,)= \eta ^{\alpha\beta\gamma}{}_\delta\,\vartheta ^ \delta\,,\label{antisym1}\\
\eta ^{\alpha\beta}&:=&\,^*(\vartheta ^\alpha\wedge\vartheta
^\beta\,)={1\over{2!}}\,\eta ^{\alpha\beta}{}_{\gamma\delta}
\,\vartheta ^\gamma\wedge\vartheta ^ \delta\,,\label{antisym2}\\
\eta ^\alpha &:=&\,^*\vartheta ^\alpha = {1\over{3!}}\,\eta ^\alpha{}_{\beta\gamma\delta}
\,\vartheta ^\beta\wedge\vartheta ^\gamma\wedge\vartheta ^ \delta\,,\label{antisym3}
\end{eqnarray}
and the four-dimensional volume element
\begin{equation}
\eta :=\,^*1 ={1\over{4!}}\,\eta _{\alpha\beta\gamma\delta}\,\vartheta ^\alpha\wedge\vartheta ^\beta\wedge\vartheta ^\gamma\wedge\vartheta ^\delta\,.\label{vol}
\end{equation}
The eta-basis features
\begin{eqnarray}
\vartheta ^\alpha\wedge\vartheta ^\beta\wedge\eta _{\alpha\beta} &=&12\,\eta\,,\label{feat1}\\
\vartheta ^\alpha\wedge\eta _\beta &=&\delta ^\alpha _\beta\,\eta\,.\label{feat2}
\end{eqnarray}
are used in the main text.

\end{document}